# A Survey on Brain-Computer Interaction

IIIT Kota

## Virtual Brain

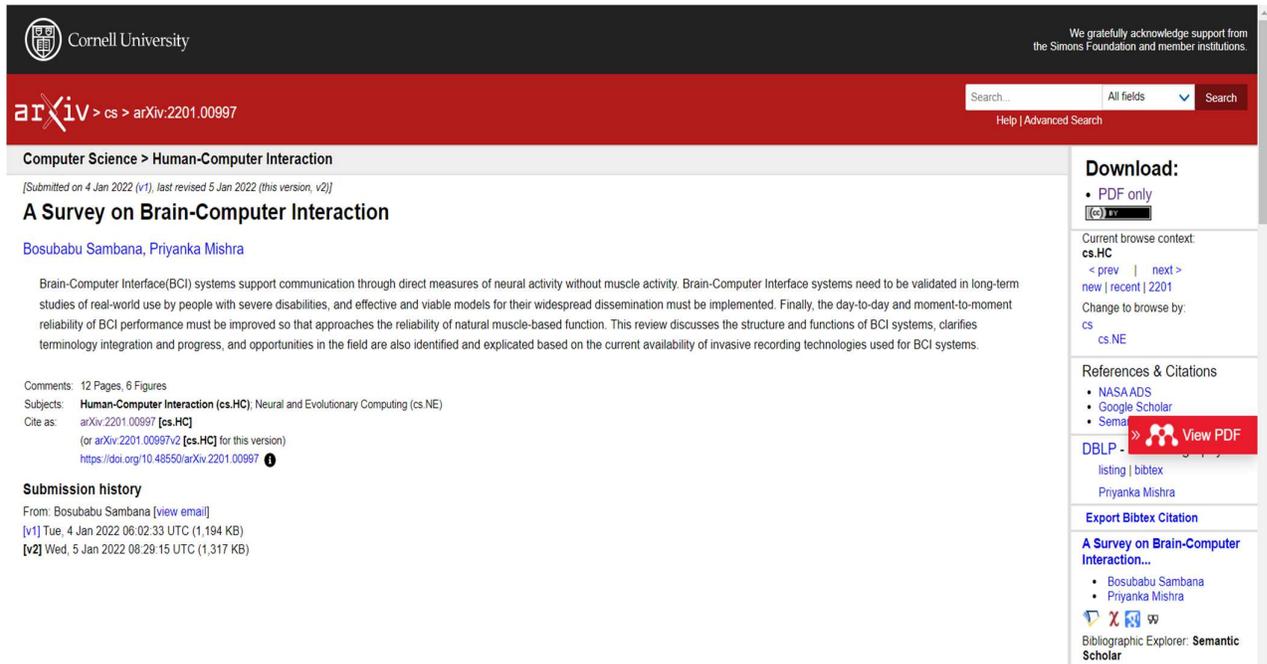

Bosubabu Sambana , Priyanka Mishra
Indian Institute of Information Technology Kota, Rajasthan, India

# A Survey on Brain-Computer Interaction

Bosubabu Sambana [1], Priyanka Mishra [2]

[1] Research Scholar, Department of Computer Science and Engineering,
Indian Institute of Information Technology Kota, Rajasthan, India.
2021krcp1001@iiitkota.ac.in

[2] Assistant Professor, Department of Computer Science and Engineering,
Indian Institute of Information Technology Kota, Rajasthan, India.
priyanka.cse@iiitkota.ac.in

**Abstract:** *Brain-Computer Interface (BCI) systems support communication through direct measures of neural activity without muscle activity. Brain-Computer Interface systems need to be validated in long-term studies of real-world use by people with severe disabilities, and practical and viable models for their widespread Dissemination must be implemented. Finally, BCI performance's day-to-day and moment-to-moment reliability must be improved to approach the reliability of natural muscle-based function. This review discusses the structure and functions of BCI systems, clarifies terminology integration and progress, and opportunities in the field are also identified and illustrated based on currently available invasive recording technologies used for BCI systems.*

## 1 Introduction

Neurological and neuroanatomical injuries and disorders affect many people worldwide and often result in movement impairment and inability to perform everyday tasks, such as communicating, reaching, and grasping, independently. Persons who have experienced neurological injuries, such as spinal cord injury (SCI), amyotrophic lateral sclerosis, or stroke, can achieve partially restored function through cortical prosthetic systems. A cortical prosthesis is an end effector device that receives an action command to perform the desired position through a brain-computer interface (BCI) that records cortical activity and extracts (i.e., decodes) information related to that intended function. End effectors can range from virtual typing communication systems to robotic arms and hands or a person's limb reanimated by functional electrical stimulation (FES).

BCI technology can range in levels of invasiveness, temporal and spatial recording resolution, and the types of recorded signals. Non-invasive brain imaging technologies, such as electroencephalography (EEG), magnetoencephalography (MEG), and functional magnetic resonance imaging (fMRI), have been implemented in various simple BCI applications, such as low throughout communication spelling systems (Speier et al., 2013). However, these non-invasive BCI options typically are too slow (e.g., fMRI), have low spatial resolution and low signal bandwidth (e.g., EEG), and are easily corrupted by external artifacts (Daly and Wolpaw, 2008; Leuthardt et al., 2004).

Thus they are not ideal for complex real-time applications, such as high-performance communication, control of multidimensional robotic limbs [1], and reanimation of paralyzed limbs for coordinated reaching and grasping. Invasive BCIs, because of the higher resolution and signal bandwidths available, have the potential to allow people with neurological injury to command more high-dimensional systems naturally and restore more complex functions.

Brain-computer interface is one of the most promising and increasingly popular technologies for assisting and improving communication/control for motor paralysis (e.g., paraplegia or quadriplegia) due to stroke, spinal cord injury, cerebral palsy, and amyotrophic lateral sclerosis (ALS). Eye-tracking technology also allows paralyzed people to control external devices. Still, it has many drawbacks due to measuring eye movements via cameras or using attached electrodes on the face, such as electrooculography (EOG) signals.

BCI essentially involves translating human brain activity into external action by sending neural commands to external devices (Belkacem et al., 2015a, 2018; Gao et al., 2017; Chen et al., 2020; Shao et al., 2020). Although the most common use of BCI is to help disabled people with disorders in the motor system, it might be a handy tool for improving the quality of life of healthy people, particularly the elderly. Assistive, adaptive, and rehabilitative BCI applications for older adults and elderly patients should be developed to assist with their domestic chores, enhance relationships with their families and improve their cognitive and motor abilities.

**BCI Platforms**
- BCI2000
- OpenViBE
- BCILAB and LSL
- TOBI Common Implementation Platform (Python PyTIAClient)
- BCI++
- BF++
- xBCI

**Related Software for Implementation**
- MNE
- SCoT
- SigViewer
- BioSig
- SNA

## 2 Related work

BCI Technology has clinical and non-clinical applications in many areas, including medicine, entertainment, education, and psychology, to solve many health issues such as cognitive deficits, slow processing speed, impaired memory, and decline in movement capability among older people. These issues can affect the quality of elderly life and may have adverse effects on mental health. Many BCI applications have been developed in the past decade to help older people maintain a healthy, good quality of life and sense of well-being.

There are two types of BCI based on the electrodes used for measuring brain activity: non-invasive BCI, where the electrodes are placed on the scalp (e.g., EEG based BCI), and invasive Brain-computer interface, where the electrodes are directly attached to the human brain [e.g., BCI based on electrocorticography (ECoG), or intracranial electroencephalography (iEEG)]. Brain-computer interfaces using EEG technology have been widely used to establish portable synchronous and asynchronous control and communication [2]. Non-invasive EEG-based BCIs can be classified as "evoked" or "spontaneous." An evoked BCI exploits a vital characteristic of the EEG, the so-called evoked potential, which reflects the immediate automatic responses of the brain to some external stimuli. Spontaneous BCIs are based on the analysis of EEG phenomena associated with various aspects of brain function related to mental tasks carried out by the BCI user at their own will.

These BCIs have been developed based on some brain features such as evoked potentials [e.g., P300 and steady-state visual evoked potential (SSVEP)] or based on slow potential shifts and variations of rhythmic activity [e.g., motor imagery (MI)]. To build a BCI system, five or six components are generally needed: signal acquisition during a specific experimental paradigm, preprocessing, feature extraction (e.g., P300 amplitude, SSVEP, or alpha/beta bands), classification (detection), translation of the classification result to commands (BCI applications), and user feedback. For quick and accurate processing and analysis of brain data, researchers have developed many open-source software packages and toolboxes such as BCI2000, EEGLab, FieldTrip, and Brainstorm.

These software packages are based on advanced signal and image processing methods and artificial intelligence programs for performing sensor or source-level analyses (Belkacem et al., 2015b, 2020; Dong et al., 2017). However, many critical issues are faced in developing a ready-to-use BCI product.

These critical issues include low classification accuracy, a small number of degrees of freedom, and a long training time to learn how to operate a BCI perfectly. Therefore, researchers have been trying to improve the performance of the existing BCIs by developing a hybrid BCI (hBCI) that combines at least two BCI modalities (e.g., P300 with SSVEP or P300 with MI).

The hBCI combines different approaches to utilize the advantages of multiple BCI modalities. It can also be a combination of brain activity with non-brain activity, and various other psychological signals were shown to be a promising option for hBCI development (Scherer et al., 2007; Choi et al., 2016). Thus, the input signals can consist of the combination of two brain characteristics using EEG signals, or EEG with eye movements (EOG), muscle activity (electromyography, EMG), or heart signal (ECG or EKG).
In addition, a closed-loop BCI system using visual and proprioceptive feedback with real-time modulation and communication can be used not only for interacting with the external environment but also as a biofeedback platform to enhance the cognitive abilities of elderly patients with better therapeutic effects [3]. This closed-loop interaction between the participant's brain responses and the stimuli is thought to induce cerebral plasticity and facilitate rehabilitation.

One of the most significant challenges in BCI technology is the development of less invasive or non-invasive technologies for paralyzed patients. Using non-invasive devices can significantly reduce both the total cost of surgical operation and the physical harm to the patient. However, non-invasive methods can lead to weaker signals and a low signal-to-noise ratio (SNR) with minor source precision and lower spatial resolution. Advanced techniques such as deep learning can partially overcome these drawbacks to decode and extract more relevant source information from the EEG signal (Nagel and Spüler, 2019).

Electroencephalogram-based BCI technology has many vital applications for motor control impairments in the medical and psychology fields. One promising application for elderly patients is the development of automatic systems to detect the influences on the brain signal related the smoking and alcohol abuse using resting-state EEG (Mumtaz et al., 2017; Su et al., 2017). BCI has also been found to help identify deficits and improve social skills in patients with autism through BCI-assisted social games (Amaral et al., 2018). Other research has focused on systems to test memory capacity and cognitive level (Burke et al., 2015; Buch et al., 2018).

The nervous system is divided into central and peripheral divisions and the separate autonomic system. The central nervous system consists of the brain and spinal cord, while the peripheral nervous system consists of the nerves to the trunk and extremities. The brain is composed of the hemispheres, the brainstem, and the cerebellum. The hemispheres are divided into the frontal, temporal, parietal, and occipital lobes: the frontal lobe houses more assertive personality and executive functions.

The central sulcus marks the dividing line between the frontal and parietal lobes, with the primary motor cortex on the anterior aspect of this sulcus and the primary sensory cortex on the posterior border.
The opercular cortex on the frontal side of the Sylvian fissure in the dominant hemisphere houses the motor speech area (Broca's area). The temporal lobe contains areas subserving memory (hippocampus), emotion and primitive urges (amygdala), hearing (primary auditory cortex, Heschel's gyrus), and speech (dominant posterior temporal region). The parietal lobe serves both primary and complex sensory functions (graphesthesia) and cortical areas for speech comprehension (Wernicke's area) and association. The occipital lobes primarily involve vision, containing primary and secondary visual regions [4].

The fibers of the corpus callosum connect the hemispheres. The primary motor and sensory cortices are organized in a homuncular pattern. The foot is in the interhemispheric fissure, and the hip area is arranged medially, with the medial-to-lateral organization of arm-hand–face–mouth.

The basal ganglia are located deep within the hemispheres and comprise the c-shaped caudate nucleus, the putamen, and the globus pallidus. Some also include the amygdala in the temporal lobe in this group of structures. Thin gray-matter bridges that connect the caudate and putamen are termed the lentiform nucleus. The anterior limb of the internal capsule passes between these bridges. The connection between these structures is more robust ventromedially. The caudate follows the curve of the ventricular system, with the tail ending in the anterior temporal lobe near the amygdala. The globus pallidus is divided into internal and external segments, the globus pallidus internal and external, respectively.

The posterior limb of the internal capsule lies at the medial border of the globus pallidus. The basal ganglia are involved in controlling both cognitive and motor functions. They form a network of deep nuclei connected to the cortex via direct and indirect pathways through the thalamus. The basal ganglia nuclei form the basis of the extrapyramidal motor control system that modulates motor function.

The brainstem is further divided into the diencephalon (thalamic complex), the mesencephalon (midbrain), the metencephalon (pons), and the myelencephalon (medulla). The diencephalon includes the thalamus, which functions as a relay center for most motor and sensory tracts (aside from olfaction), the epithalamus (pineal, habenular nuclei, stria medullaris), which functions to control the daily cycle, the hypothalamus, which regulates multiple pituitary hormones via release factors as well as producing its hormones (ADH, vasopressin), and the subthalamic nucleus, which participates in the extrapyramidal motor system along with the basal ganglia. The thalamus is bounded laterally by the posterior limb of the internal capsule and caudally by the midbrain. The ventrocaudal($V_c$) nucleus (sometimes referred to as the ventral posterior or VP nucleus) is primarily responsible for relaying painful sensations.

The midbrain contains the nuclei for cranial nerves III and IV and the corticobulbar and corticospinal tracts carrying motor fibers from the cortex to the brainstem and spinal cord, respectively (see in the cerebral peduncles). The fibers of the dentatorubrothalamic tract decussate in the midbrain after emerging from the superior cerebellar peduncle on their way to the red nucleus in the midbrain tegmentum and then on to the thalamus. The midbrain's dorsal aspect (tectum) consists of the paired superior and inferior colliculi subserving coordination of vision and hearing, respectively. The pedunculopontine nucleus is also located here. This nucleus functions in the extrapyramidal motor system as part of the "locomotor center." It has connections to the pallidum, cortex, and substantia nigra.

A brain-computer interface (BCI) is a communication system by which a person can send messages or commands without any use of peripheral nerves and muscles. BCIs record signals from the brain and translates them into helpful communication. Thus, they are usable even by people who have no effective muscle control. This review describes the essential components and the major categories of current BCIs, defines terms used in the BCI literature, and considers advances that might be expected in the next few years.

Brain-computer interfaces (BCIs) acquire brain signals, analyze them, and translate them into commands relayed to output devices that carry out desired actions. BCIs do not use normal neuromuscular output pathways. The main goal of BCI is to replace or restore helpful function to people disabled by neuromuscular disorders such as amyotrophic lateral sclerosis, cerebral palsy, stroke, or spinal cord injury.

From initial demonstrations of electroencephalography-based spelling and single-neuron-based device control, researchers have used electroencephalographic, intracortical, electrocorticographic, and other brain signals for increasingly complex control of cursors, robotic arms, prostheses, wheelchairs, and other devices. Brain-computer interfaces may also prove helpful for rehabilitation after stroke and other disorders.

In the future, they might augment the performance of surgeons or other medical professionals. Brain-computer interface technology focuses on a rapidly growing research and development enterprise that is greatly exciting for scientists, engineers, clinicians, and the public.

Its future achievements will depend on advances in 3 crucial areas. Brain-computer interfaces need signal-acquisition hardware that is convenient, portable, safe, and able to function in all environments. Brain-computer interface systems need to be validated in long-term studies of real-world use by people with severe disabilities, and practical and viable models for their widespread Dissemination must be implemented. Finally, BCI performance's day-to-day and moment-to-moment reliability must be improved to approach the reliability of natural muscle-based function.

Until recently, the dream of being able to control one's environment through thoughts had been in the realm of science fiction. However, advances in technology have brought a new reality: Today, humans can use the electrical signals from brain activity to interact with, influence, or change their environments. The emerging field of brain-computer interface (BCI) technology may allow individuals unable to speak and use their limbs to once again communicate or operate assistive devices for walking and manipulating objects.

Brain-computer interface research is an area of high public awareness. Videos on YouTube and news reports in the lay media indicate intense curiosity and interest in a field that hopefully, one day soon, will dramatically improve the lives of many disabled persons affected by several different disease processes. This review seeks to provide the general medical community with an introduction to BCIs.

We define BCI and then review some of the seminal discoveries in this rapidly emerging field, the brain signals used by BCIs, the essential components of a BCI system, current BCI systems, and the key issues now engaging researchers. Challenges are inherent in translating any new technology to practical and valuable clinical applications, and BCIs are no exception.

We discuss BCI systems' potential uses and users and address some of the field's limitations and challenges.

We also consider the advances that may be possible in the next several years. Recently published a detailed presentation of BCI technology's basic principles, current state, and prospects [4].

A BCI is a computer-based system that acquires brain signals, analyses them, and translates them into commands relayed to an output device to carry out the desired action. Thus, BCIs do not use the brain's normal output pathways of peripheral nerves and muscles.

This definition strictly limits the term BCI to systems that measure and use signals produced by the central nervous system (CNS). Thus, for example, a voice-activated or muscle-activated communication system is not a BCI. Furthermore, an electroencephalogram (EEG) machine alone is not a BCI because it only records brain signals but does not generate an output that acts on the user's environment.

It is a misconception that BCIs are mind-reading devices. Brain-computer interfaces do not read minds to extract information from unsuspecting or unwilling users but enable users to act on the world using brain signals rather than muscles. The user and the BCI work together. After training, the user generates brain signals that encode intention. After training, the BCI decodes the signals and translates them into commands to an output device that accomplishes the user's preference.

Brain-computer interfacing is an emerging technology that connects a brain with external devices, providing a new output channel for brain signals to communicate with or control such devices without using natural neuromuscular pathways.

A brain-computer interface (BCI) recognizes the user's intent through brain signals, decodes neural activity, and translates it into output commands that accomplish the user's goal. BCI technology has the potential to restore lost or impaired functions of people severely disabled by various devastating neuromuscular disorders or spinal cord damage and to enhance or augment tasks in healthy individuals. Different brain signals have decoded user intent in BCI research, ranging from direct neuronal recordings using implanted electrodes to non-invasive recordings such as scalp electroencephalogram (EEG).

Donoghue provides a critical review of the state of the art of invasive intracortical BCIs based on neural recordings obtained using implanted electrodes. He summarizes where the field is now regarding the neuroscience and engineering challenges that remain before invasive BCIs become practical and generally available. Supporting discussion asserts that fully implantable microelectronics systems capable of signal processing and wireless transmission and devices for high-throughput generation of command signals are complex engineering challenges that will play a critically important role in future advancements.

Bouton reviews essential issues related to invasive BCIs and a novel approach for restoring motor functions in a human subject by combining an invasive BCI with functional electrical stimulation. This approach extends the state of art from a BCI-controlled robotic arm to a BCI-controlled functional restoration of arm movement in a paralyzed person.

Ajiboye and Kirsch discuss components of invasive BCI systems, including the advantages and disadvantages of various recording technologies, potential cortical areas of implantation, signals of interest, and how these signals are decoded into operational commands.

Shoffstall and Capadona discuss the motivation, progress, challenges, and prospects for chronically stable intracortical recording electrodes for invasive BCI applications. Schalk and Allison discuss non-invasive BCI techniques using various measurement modalities, such as EEG, for decoding a user's intent. Multiple applications are discussed, including replacing lost functions, restoring the ability to control the body, improving or enhancing operations, and augmenting the body's natural outputs.

- Brain-computer interfaces (BCIs) are provided for people with severe disabilities to use in their homes.
- Effective information throughput is being improved by developing or enhancing sensor and hardware technology; signal processing and translation approaches; error correction and response verification; word and sentence selection and/or completion algorithms; additional signals, including hybrid BCIs sequential menus; and goal-oriented protocols.
- The right BCI for a given user can be found by considering factors including performance, fatigue, training time, invasiveness, reliability, cost, flexibility, environment, cosmesis, comfort, are of set-up and use, the user's needs, desires, motivation abilities, and access to assistance with preparing, using, repairing, cleaning or updating the BCI.
- BCIs can be integrated with conventional computers, medical equipment, headwear, software, accessories, and interfaces, allowing more flexible, usable mainstream BCIs.
- BCI-related clinical and research infrastructure should continue to be improved to provide information to and among

researchers, medical personnel, patients, other user-supporting staff, students, potential and actual funding sources, the media, and the public.

**Signal Sensor System**

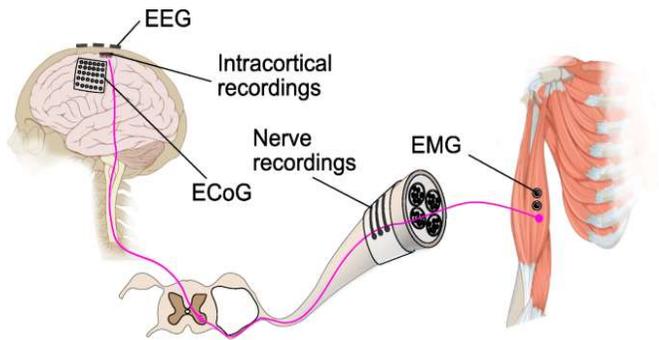

Signals that might be used for BCIs can be recorded from four locations as shown schematically in from sensors that are not in contact with the body, such as in functional MRI (fMRI) or magnetoencephalography (MEG); from the surface of the scalp via standard electroencephalographic (EEG) electrodes or functional near-infrared (fNIR) spectroscopy; from the surface of the dura or the surface of the brain using ElectroCorticoGraphic (ECoG) electrodes; or from within the brain using microelectrodes implanted in the cortex or elsewhere in the brain. In healthy and severely disabled people, signals from these areas can be extracted and translated for communication and control.

The basic structure of any BCI. A BCI has four essential components: the signal acquisition component, which records brain signals at one of the sites described above; the signal processing component, which includes the software that extracts the features of the brain signals that are used for the BCI and the translation algorithm that translates the extracted features into device commands; the output device component that implements the controls; and the operating protocol that governs how these components interact.

**Autonomic Nervous System**

The autonomic nervous system innervates the glands, viscera, heart, and smooth muscle. This system is divided into parasympathetic and sympathetic divisions. Each division consists of ganglia with both preganglionic and postganglionic branches. The sympathetic ganglia are located in either the paraspinal chain or the prevertebral plexuses. The sympathetic system originates in the posterior hypothalamus and medulla.

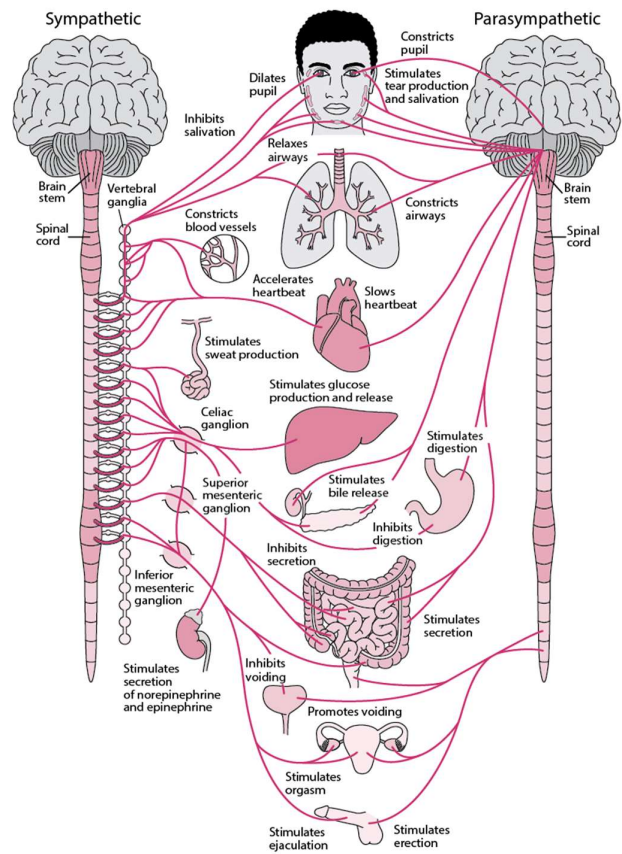

Efferent (fibers leaving a nucleus and traveling to a target) journey into the spinal cord, where they synapse in the intermediolateral cell column, located between T2 and L1. The preganglionic myelinated fibers travel as white rami communicate to the paravertebral chain and either synapse and send post-ganglionic gray (unmyelinated) rami corresponding to the spinal nerves or pass through to the prevertebral ganglia that supply autonomic innervation to the viscera. The sympathetic fibers to the head extending from the superior cervical ganglion to follow the external carotid artery branches.

**Sensory System**

First-order afferents for delicate touch and position sense have their nuclei in the dorsal root ganglia and travel centrally in the dorsal columns (fasciculus cuneatus and fasciculus gracilis) of the spinal cord, synapsing in the medullary dorsal column nuclei (nucleus cuneatus and nucleus gracilis). They then cross and ascend as the medial lemniscus to the $V_c$ nucleus of the thalamus. From there, third-order fibers travel to the sensory cortex. The thalamic homunculus is arranged such that the face is medial, the lower body is lateral, and the upper extremity lies between.

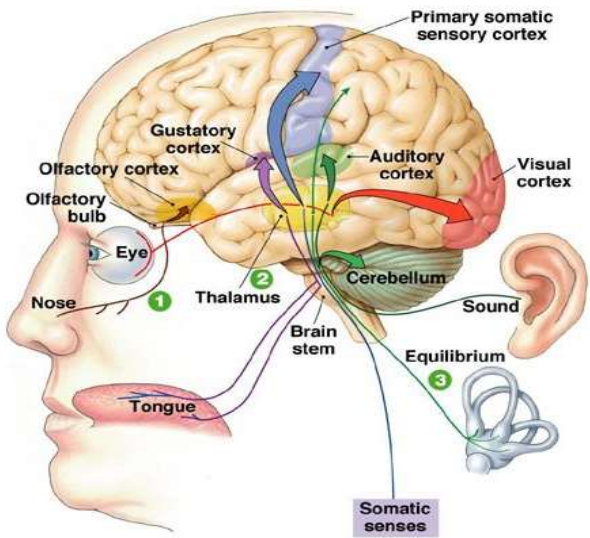

Central processes of afferents for pain and temperature extend from the dorsal root ganglia to synapse in Lissauer's tract (as described below) and then cross in the cord just ventral to the central canal to form the contralateral ventral spinothalamic tract, with fibers organized with the lower body more laterally and the upper body more medially. These ascend to the $V_c$ thalamus and send third-order neurons to the sensory cortex from the thalamus.

Rexed first described the laminar organization of the spinal gray matter in the 1950s. Afferent fibers enter the DH via the dorsolateral fasciculus of Lissauer. Afferent spinothalamic axons may travel vertically over several spinal segments in this superficial layer before eventually synapsing with neurons in lamina I; the posted marginal nucleus.

This layer contains nociceptive-specific neurons that respond almost exclusively to noxious stimuli (Byers and Bonica, 2001; Carpenter, 1991c; Terman and Bonica, 2001). They include multiple neuropeptides, including substance P, calcitonin gene-related peptide (CGRP), enkephalin, and serotonin. Substance P and CGRP, in particular, play an important role in DH nociception (Donnerer and Amann, 1992; Donnerer et al., 1992a,b; Donnerer and Stein, 1992). Lamina I cells send axons contralaterally across the ventral aspect of the central canal to form the lateral spinothalamic tract (STT). Lamina I also contains a class of cells that respond to a large variety of both noxious and nonnoxious stimuli.

This wide dynamic range (WDR) cells can substantially alter their discharge frequency to reflect the input stimulus type. Noxious stimuli evoke higher frequency discharges from WDR cells [3]. As described below,

Bosubabu Sambana , Priyanka Mishra

these cells play an essential role in developing chronic neuropathic pain.

Lamina II, the substantial gelatinosa, modulates input from sensory receptors. Nociceptive and chemoreceptive input is concentrated in the superficial layer of this lamina ($II_o$), while mechanoreceptor input is targeted to the more profound aspect ($II_i$) (Carpenter, 1991c; Terman and Bonica, 2001). Projections from substantia gelatinosa neurons terminate in lamina I and lamina II at other spinal levels. Opiate receptors are plentiful in both laminae I and II. Importantly, each sublayer of lamina II appears to contain distinct subpopulations of C fibers. Those C fibers terminating in lamina II are similar to those completing in lamina I, and they express substance P and CGRP and contain the tyrosine kinase (trk) A receptor for nerve growth factor.

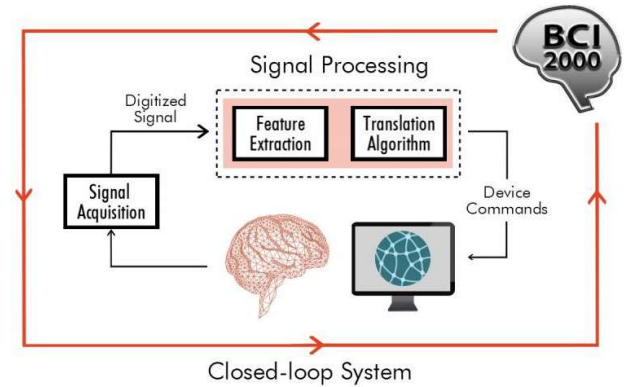

In contrast, the C fibers terminating in lamina $II_i$ do not express either CGRP or substance P and communicate the binding site for lectin IB4, an indicator of sensitivity to glial-derived neurotrophic factor. This lamina also contains numerous local circuit neurons whose dendritic arbors may extend into more deep and superficial laminae.

The A-β fibers terminate primarily in lamina III, as do some of the A-δ mechanoreceptive fibers. Lamina IV also serves as a target zone for A-β fibers. Some of the cells in this layered project back to layer I, aiding in integrating sensory information. Lamina V contains many STT projection cells that receive input from A-δ and C fibers. A substantial proportion of the cells here are WDR neurons. These have large receptive fields whose centers are responsive to both noxious and noxious stimuli and surrounding areas responsive primarily to noxious stimuli only. Stimulation of the region surrounding this field causes inhibition of the WDR neuron (Terman and Bonica, 2001).

## BCI Signal Acquisition

Most BCIs do not require surgery to implant electrodes and are termed non-invasive BCIs. At present, almost all non-invasive BCIs measure brain activity with EEG sensors placed on the scalp's surface; this review focuses mainly on such BCIs. BCIs that acquire signals from electrodes surgically implanted in or on the cortex or other brain areas are invasive. ECoG-based BCIs are invasive because they require surgery but are less invasive than intracortical BCIs since ECoG electrodes do not penetrate the brain but lie on the brain's surface.

Invasive electrodes may give a more detailed view of brain activity than non-invasive systems. Since the scalp smears, dampens and filters the brain's electrical activity, invasive electrodes may allow better spatial resolution, stronger signals, and a more comprehensive range of frequencies than electrodes placed on the scalp. For example, ECoG BCIs can detect movement-related activity in the 100–200-Hz range, well beyond the scope of scalp electrodes.

Invasive BCIs can be available 24 h per day, require less preparation and clean-up time, and are less susceptible to noise from muscle artifacts and external noise. However, invasive BCIs currently offer approximately the same performance as non-invasive systems. Furthermore, they entail expensive surgery, scarring, risk of infection, and regular medical check-ups, and their long-term stability remains unclear. Hence, while invasive BCIs merit further study, most patients and researchers may, understandably, choose non-invasive approaches.

## BCI output devices

After the brain signal features are extracted and translated, the third component of the BCI, the output device, implements the messages or commands conveyed by the translation algorithm. To date, the most commonly used BCI out-put device is a computer monitor. Monitor-based BCIs have been developed in which users move a cursor to chosen targets in one or more dimensions, select one item from two or more choices, select items from a scrolling or iterative menu, browse the internet or navigate a virtual environment.

Some BCIs use speakers or headphones to provide auditory stimulation or feedback. BCIs have also been used to control switches, common appliances, such as an air conditioners, television or music player, medical devices, robotic arms, mobile robots, functional electrical stimulators or orthoses, and a full-motion flight simulator.

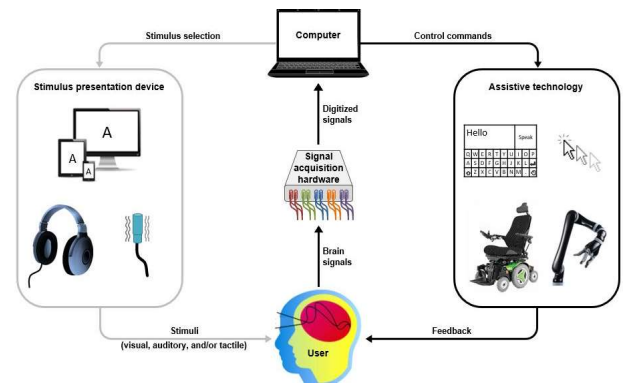

## BCI operating protocols

The operating protocol defines the real-time interactions between the user's brain and the BCI system. It provides a front end for the user and operator, governs how the other three modules interact with each other and the operating system, and mediates details of user-system interaction, such as what selections are available to the user, when and how user activity may affect control and the nature and timing of feedback. Similar to other BCI components, operating protocols have advanced significantly in several years. Many papers have addressed timing, feedback, and usability.

Error correction based on EEG activity, such as the error-related negativity, P300, or other measures, may improve performance in some users. Some operating protocols allow a much larger vocabulary than most early BCIs, either by presenting many options or letting the user select from among different palettes of options, sometimes via a menu.

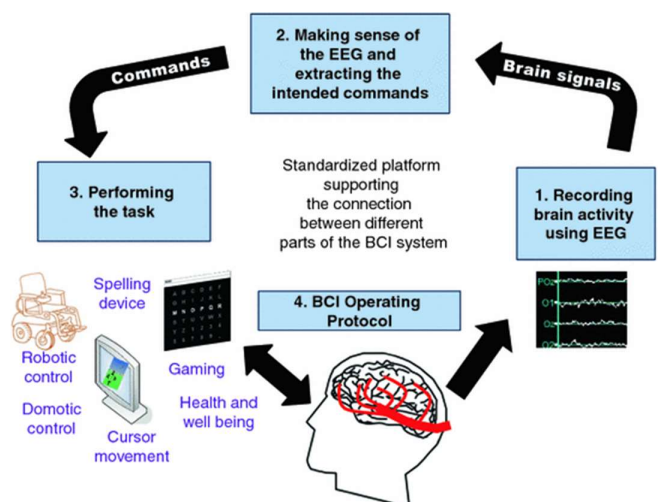

The most widely used BCI operating system is BCI2000, which over 150 laboratories worldwide are currently being used. BCI2000 has many features that make it appealing to researchers, and it is highly flexible and interchangeable. It has been validated with the various signal acquisition, signal processing, and output systems. It offers a variety of real-time and offline analyses and is available free of charge for research use.

Feature Extraction Feature extraction analyzes the digital signals to distinguish pertinent signal characteristics (i.e., signal features related to the person's intent) from irrelevant content. It represents them in a compact form suitable for translation into output commands. These features should have strong correlations with the user's intent. Because much of the relevant (i.e., most strongly correlated) brain activity is either transient or oscillatory, the most commonly extracted signal features in current BCI systems are time-triggered EEG or ECoG response amplitudes and latencies, power within specific EEG or ECoG frequency bands, or firing rates of individual cortical neurons.

Environmental and physiologic artifacts such as electromyographic signals are avoided or removed to ensure accurate measurement of the brain signal features.

Feature Translation The resulting signal features are then passed to the feature translation algorithm, which converts the components into the appropriate commands for the output device (i.e., commands that accomplish the user's intent). For example, a power decrease in a given frequency band could be translated into an upward displacement of a computer cursor, or a P300 potential could be translated into selecting the letter that evoked it.

The translation algorithm should be dynamic to accommodate and adapt to spontaneous or learned changes in the signal features and ensure that the user's possible range of feature values covers the full range of device control.

### The Future of BCIs: Problems and Prospects

Brain-computer interface research and development generates tremendous excitement in scientists, engineers, clinicians, and the general public. This excitement reflects the rich promise of BCIs. They may eventually be used routinely to replace or restore helpful function for people severely disabled by neuromuscular disorders; they might also improve rehabilitation for people with strokes, head trauma, and other conditions.

At the same time, this exciting future can come about only if BCI researchers and developers engage and solve problems in 3 critical areas: signal-acquisition hardware, BCI validation and Dissemination, and reliability. Signal-Acquisition Hardware All BCI systems depend on the sensors and associated hardware that acquire the brain signals. Improvements in this hardware are critical to the future of BCIs.

Ideally, EEG-based (non-invasive) BCIs should have electrodes that do not require skin abrasion or conductive gel (i.e., so-called dry electrodes); be small and fully portable; have comfortable, convenient, and cosmetically acceptable mountings; be easy to set up; function for many hours without maintenance; perform well in all environments; operate by telemetry instead of requiring wiring; and interface easily with a wide range of applications.

In principle, many of these needs could be met with current technology, and dry electrode options are becoming available (e.g., from g.tec Medical Engineering, Schiedlberg, Austria). The achievement of good performance in all environments may prove the most challenging requirement. Brain-computer interfaces that use implanted electrodes face a range of complex issues. These systems need hardware that is safe and fully implantable; remains intact, functional, and reliable for decades; records stable signals over many years; conveys the recorded calls by telemetry; can be recharged in situ (or has batteries that last for years or decades); has external elements that are robust, comfortable, convenient, and discreet; and interfaces easily with high-performance applications [2]. Although great strides have been made in recent years, and individual cases of microelectrode implants have continued to function over the years, it is unclear which solutions will be most successful.

ECoG- or local field potential-based BCIs might provide more consistently stable performance than BCIs that rely on neuronal action potentials. Nevertheless, significant as yet undefined innovations in sensor technology may be required for invasive BCIs to realize their full promise. Much of the necessary research will continue to rely primarily on animal studies before the initiation of human trials.

Validation and Dissemination As work progresses and BCIs begin to enter actual clinical use, two essential questions arise: how good a given BCI can get (e.g., how capable and reliable) and which BCIs are best for which purposes[1]. To answer the first question, each promising BCI should be optimized, and the limits on users' capabilities should be defined. Addressing the second question will require consensus among research groups regarding which applications should be used for comparing BCIs and how performance should be assessed.

The most obvious example is whether the performance of BCIs that use intracortical signals is significantly superior to that of BCIs that use ECoG signals or even EEG signals. For many prospective users, invasive BCIs will need to provide much better performance to be preferable to non-invasive BCIs, and it is not yet certain that they can do so. The data to date do not answer this critical question.126 On the one hand, it may turn out that non-invasive EEG- or fNIR-based BCIs are used primarily for essential communication, while ECoG- or neuron-based BCIs are used for complex movement control.

On the other hand, non-invasive BCIs may prove nearly or equally capable of complex uses. At the same time, invasive BCIs that are fully implantable (and thus very convenient to use) might be preferred by some people, even for primary communication purposes. At this point, many different outcomes are possible, and the studies and discussions necessary to select among them have just begun.

The development of BCIs for people with disabilities requires clear validation of their real-life value in efficacy, practicality (including cost-effectiveness), and impact on quality of life. This depends on multidisciplinary groups being able and willing to undertake lengthy studies of real-life use in complicated and often challenging environments. Such studies, which are just beginning (e.g., by Sellers et al. 41), are an essential step if BCIs are to realize their promise.

The validation of BCIs for rehabilitation after strokes or in other disorders will also be demanding and require careful comparisons with the results of conventional methods alone. Current BCIs, with their limited capabilities, are potentially helpful, mainly for people with very severe disabilities.

Bosubabu Sambana , Priyanka Mishra

Because this user population is relatively small, these BCIs are essentially orphan technology, and there is not yet adequate incentive for commercial interests to produce them or promote their widespread Dissemination. Invasive BCIs entail substantial costs for initial implantation, plus the cost of ongoing technical support.

Although the initial costs of non-invasive BCI systems are relatively modest (for example, $5,000-$10,000), they too require some measure of ongoing technical support. The future commercial practicality of all BCIs will depend on reducing the amount and sophistication of the long-term support needed, increasing the numbers of users, and on ensuring reimbursement from insurance companies and government agencies. Clear evidence that BCIs can improve motor rehabilitation could significantly increase the potential user population.

In any case, if and when further work improves the functionality of BCIs and renders them commercially attractive, their Dissemination will require viable business models that give both financial incentives for the commercial company and adequate reimbursement to the clinical and technical personnel who will deploy and support the BCIs. The optimal scenario could be in which BCIs for people with severe disabilities develop synergistically with BCIs for the general population.

### BCI versus BMI

While the term BCI now predominates in scientific and popular literature, other words are sometimes used to describe a BCI system. These include: brain-machine interface (BMI), direct brain interface (DBI), brain interface, cognitive neural prosthetics, neural interface system, and brain actuated control. Although efforts are sometimes made to give these terms different meanings, they all mean essentially the same thing: a communication and control system that uses signals generated in the CNS and does not depend on peripheral nerves or muscles.

### Devices that are not BCIs

Some devices record brain signals but are not BCIs. For example, devices that evaluate cognitive and neural activity associated with alertness or workload, sleep stage, sleep apnea, depth of anesthesia, deception, error detection, or image recognition may seem similar to BCIs but do not provide the user with real-time communication or control. Systems that send signals to the brain are not BCIs, though they might be called Computer–Brain interfaces (CBIs).

Finally, it is essential to distinguish actual BCIs from systems that use non-CNS signals recorded from the head, such as EMG or electrooculographic activity.

## Components of BCI System

A brain-computer interface (BCI) is a computer-based system that acquires brain signals, analyses them, and translates them into commands relayed to an output device to carry out the desired action.

In principle, any type of brain signal could be used to control a BCI system. The most commonly studied signals are electrical signals from brain activity measured from electrodes on the scalp, the cortical surface, or the cortex.

The purpose of a BCI is to detect and quantify features of brain signals that indicate the user's intentions and to translate these features in real-time into device commands that accomplish the user's intent.

A BCI system consists of Four (4) sequential components:

1) Signal Acquisition,
2) Feature Extraction,
3) Feature Translation,
4) Device Output.

These four components are controlled by an operating protocol that defines the onset and timing of the operation, the details of signal processing, the nature of the device commands, and the oversight of performance. An effective operating protocol allows a BCI system to be flexible and serve each user's specific needs.

At present, the striking achievements of BCI research and development remain confined almost entirely to the research laboratory. Studies that seek to demonstrate BCI practicality and efficacy for long-term home use by people with disabilities are just beginning.

Brain-computer interfaces may eventually be used routinely to replace or restore helpful functions for people severely disabled by neuromuscular disorders and augment natural motor outputs for pilots, surgeons, and other highly skilled professionals. Brain-computer interfaces might also improve rehabilitation for people with strokes, head trauma, and other disorders.

The future of BCIs depends on progress in 3 critical areas:

1). Development Of Comfortable,
2). Convenient, And Stable Signal-Acquisition Hardware;
3). BCI validation, Dissemination, and proven BCI reliability and value for many different user populations.

## Milestones in BCI Development

Can observable electrical brain signals be put to work as carriers of information in person-computer communication or to control devices such as prostheses? That was the question posed by Vidal in 1973.2 His Brain-Computer Interface Project was an early attempt to evaluate the feasibility of using neuronal signals in a personal computer dialogue that enabled computers to be a prosthetic extension of the brain. Although work with monkeys in the late 1960s showed that signals from single cortical neurons could be used to control a meter needle, systematic investigations with humans began in the 1970s.

Initial progress in human BCI research was slow and limited by computer capabilities and our knowledge of brain physiology. By 1980, Elbert et al. demonstrated that persons given biofeedback sessions of slow cortical potentials in EEG activity could change those potentials to control the vertical movements of a rocket image traveling across a television screen.

In 1988, Farwell and Donchin showed how the P300 event-related potential could allow regular volunteers to spell words on a computer screen. Since the 1950s, the mu and beta rhythms (i.e., sensorimotor rhythms) recorded over the sensorimotor cortex were associated with movement or movement imagery.6 In the late 1970s, Kuhlman7 showed that the mu rhythm could be enhanced by EEG feedback training.

Starting from this information, Wolpaw et al. trained volunteers to control sensorimotor rhythm amplitudes and use them to move a cursor on a computer screen accurately in 1 or 2 dimensions. By 2006, a microelectrode array was implanted in the primary motor cortex of a young man with complete tetraplegia after a C3-C4 cervical injury. Using the signals obtained from this electrode array, a BCI system enabled the patient to open the simulated e-mail, operate a television, open and close a prosthetic hand, and perform rudimentary actions with a robotic arm.

In 2011, Krusienski and Shih demonstrated that signals recorded directly from the cortical surface (electrocorticography [ECoG]) could be translated by a BCI to allow a person to spell words on a computer screen accurately. Brain-computer interface research is multiplying, as evidenced by the number of peer-reviewed publications in this field over the past ten years.

## 3  Conclusion

Many researchers worldwide are developing BCI systems that were in the realm of science fiction a few years ago. These systems use different brain signals, recording methods, and signal-processing algorithms. They can operate many other devices, from cursors on computer screens to wheelchairs to robotic arms. A few people with severe disabilities already use a BCI for essential communication and control in their daily lives. With better signal-acquisition hardware, clear clinical validation, viable dissemination models, and, probably most importantly, increased reliability, BCIs may become an important new communication and control technology for people with disabilities and possibly the general population.